\documentclass[11pt]{article}

\usepackage[utf8]{inputenc}
\usepackage[T1]{fontenc}
\IfFileExists{lmodern.sty}{\usepackage{lmodern}}{}
\usepackage[margin=1in]{geometry}
\usepackage{amsmath,amssymb,amsfonts}
\usepackage{graphicx}
\usepackage{float}
\usepackage{authblk}
\usepackage{caption}
\usepackage{setspace}
\usepackage{textcomp}
\IfFileExists{lmodern.sty}{\usepackage{microtype}}{\usepackage[expansion=false]{microtype}}
\usepackage[hidelinks,breaklinks=true]{hyperref}
\usepackage{url}

\title{\textbf{One-for-All Adaptive Radiotherapy Planning Agent:\\
A Foundation Framework for Daily CBCT-guided Radiotherapy}}

\author[1,2,3]{Shaoyan Pan}
\author[1]{Kirk Jon Luca}
\author[1]{Yuan Gao}
\author[3]{Shansong Wang}
\author[1,2]{Mingzhe Hu}
\author[1]{Ryan Sanford}
\author[3]{Mojtaba Safari}
\author[1]{Justin Roper}
\author[3]{Zhen Tian}
\author[4]{Tonghe Wang}
\author[1,2,3]{Xiaofeng Yang\thanks{Corresponding author: Xiaofeng Yang (\href{mailto:xfyang@uchicago.edu}{xfyang@uchicago.edu})}}

\affil[1]{Department of Radiation Oncology and Winship Cancer Institute, Emory University, Atlanta, GA 30322, USA}
\affil[2]{Department of Biomedical Informatics, Emory University, Atlanta, GA 30322, USA}
\affil[3]{Department of Radiation and Cellular Oncology, The University of Chicago, Chicago, IL 60637, USA}
\affil[4]{Department of Medical Physics, Memorial Sloan Kettering Cancer Center, New York, NY 10065, USA}

\date{}

\begin{document}

\maketitle


\begin{abstract}
Adaptive radiotherapy treatment planning seeks to enable accurate and individualized cancer care by ensuring that radiation dose conforms to the target while minimizing exposure to surrounding healthy tissues. Its success depends on accurate characterization of patient-specific anatomy at the time of treatment, whereas standard radiotherapy planning generally uses one pretreatment CT dataset for the entire treatment course, despite ongoing changes in disease extent and internal organ position. While daily adaptive radiotherapy (ART) can mitigate these discrepancies, it remains clinically impractical as each patient and fraction requires a bespoke, labor-intensive workflow. This process necessitates complex clinical decision-making---ranging from simple registration to a complete plan redesign involving CT collection, re-segmentation, and dose optimization---creating an unsustainable burden on hospital resources. In this work, we introduce the One-for-All Adaptive Radiotherapy Planning Agent, a unified foundation-model-based system that performs complete, treatment-specific online adaptive planning directly from daily cone-beam CT in under two minutes. The agent first autonomously predicts all essential planning components, including synthetic CT generation, multimodal alignment, and tumor/organ segmentation. It then intelligently leverages these outputs to execute the final clinical plan design, providing a comprehensive, automated solution for daily treatment. We also demonstrate that the agent enables clinicians to define planning with intent and intervene at critical decision points, ensuring a ``human-in-the-loop'' framework that generates acceptable plans before final approval. Evaluated on multiple datasets spanning head-and-neck, lung, abdominal, and prostate cancers with both photon and proton therapy, the proposed framework achieves clinically acceptable accuracy and plan quality comparable to clinically generated treatment plans, with target dose errors ($D_{98}$) generally within 2.0~Gy of the reference plan. The strong performance of the One-for-All agent highlights the promise of a unified foundation-model approach and opens opportunities for fast, scalable, and fully automated online adaptive radiotherapy across diverse clinical scenarios.
\end{abstract}

\section*{Introduction}

Radiotherapy is a cornerstone of cancer treatment\cite{sung2021} and is delivered to more than half of patients with cancer \cite{barton2014,ahmad2019,deruysscher2019}, contributing substantially to curative care\cite{atun2015}. Its effectiveness depends on accurate treatment planning to ensure adequate dose delivery to the tumor while minimizing irradiation of surrounding organs-at-risk (OARs)\cite{goitein1992,marks2010}. In current clinical practice, however, radiotherapy planning is largely based on an initial simulation computed tomography (CT) scan acquired before treatment begins\cite{jaffray2012,beaton2019}. At the time of treatment and throughout the treatment course, however, patient anatomy may have substantially changed due to tumor regression or progression, local organ deformation, weight loss, or other internal variations\cite{bobic2023,langen2001,barker2004,vanderhorst2017}. As a result, plans derived from an initial CT scan may no longer accurately reflect the anatomy at the time of dose delivery\cite{hammingvrieze2021,jafar2026}, potentially compromising tumor coverage and increasing normal tissue toxicity\cite{schwartz2012,bryant2024}.

\begin{figure}[H]
\centering
\includegraphics[width=0.85\textwidth]{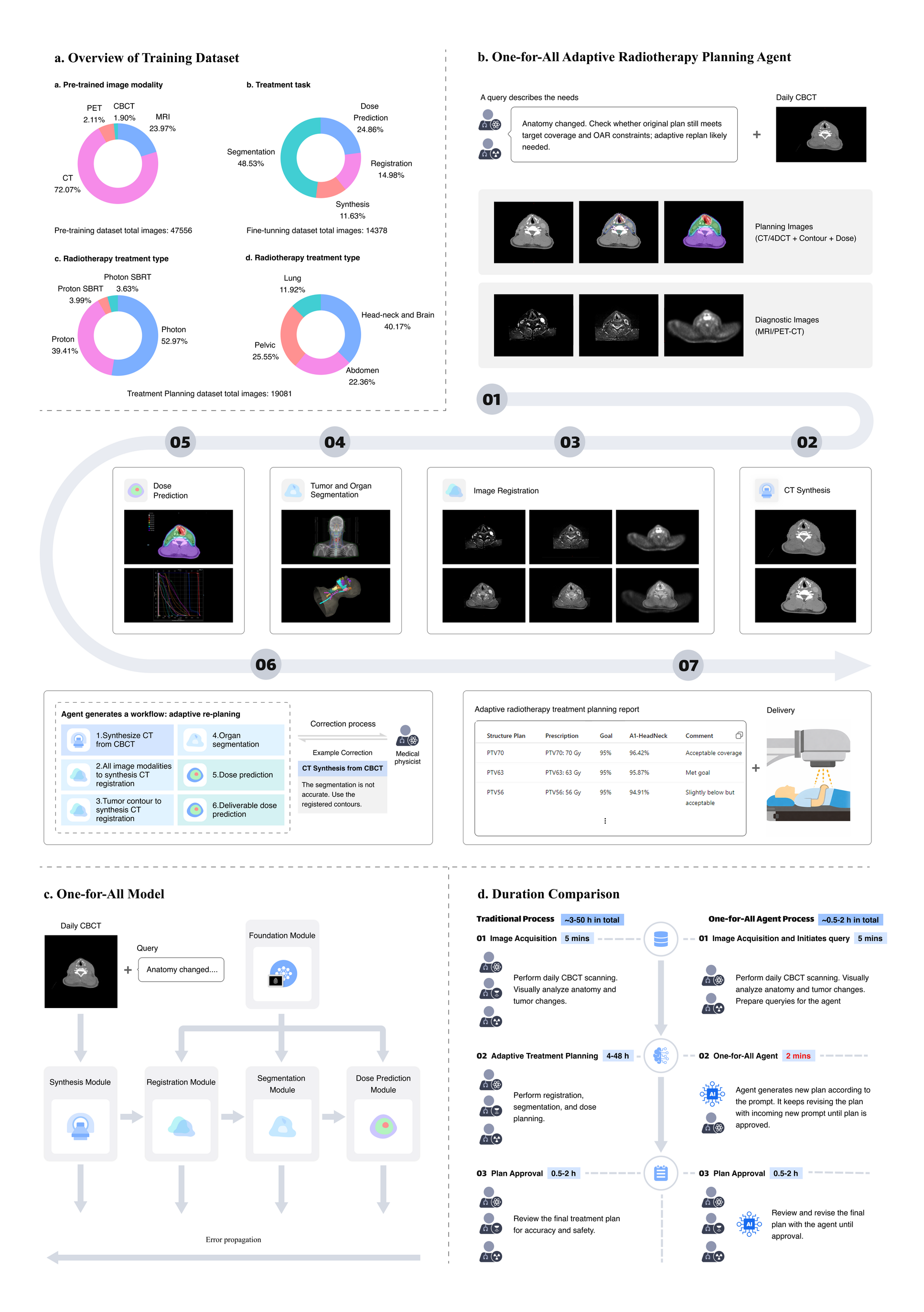}
\caption{\textbf{Overview of the One-for-All adaptive radiotherapy planning framework.} \textbf{a,} Composition of the training dataset used to develop the unified radiotherapy foundation model, showing the distributions of image modalities used for pre-training and the downstream treatment-planning tasks used for fine-tuning. The panel also indicates radiotherapy delivery techniques and anatomical treatment sites supported by the A-RPA framework in this study. \textbf{b,} Clinical workflow of the One-for-All Adaptive Radiotherapy Planning Agent. Given a clinician query and the daily on-board CBCT acquired immediately before treatment, the agent integrates prior planning data, including planning CT or 4DCT, contours and prior dose, together with available diagnostic imaging, such as MRI or PET-CT, and performs CT synthesis, multimodal image registration, tumor and OAR segmentation, and dose prediction to generate an adaptive replanning workflow for clinician review before treatment delivery. \textbf{c,} Architecture of the unified One-for-All model. The synthesis module first takes the daily CBCT and clinician query to generate a planning-ready synthetic CT and planning-aware features. The synthetic CT and the same query are then used as conditioning signals for the shared foundation module, which supports the registration, segmentation and dose-prediction modules and enables coordinated downstream inference within a unified adaptive planning framework. \textbf{d,} Comparison of the conventional adaptive radiotherapy workflow with the proposed agent-based framework. Whereas the standard workflow requires sequential expert-driven steps for adaptive replanning over several hours, the proposed agent generates a treatment-specific adaptive plan in approximately 2 minutes, which is then ready for final clinician approval.}
\label{fig:overview}
\end{figure}

Adaptive radiotherapy \cite{bryant2024,keall2022,piperdi2021,donalemus2024} addresses this limitation directly by modifying the treatment plan in response to the patient's anatomy at a given treatment fraction. In principle, this approach offers the theoretical advantage of preserving target coverage and organ sparing despite ongoing anatomical change. In practice, however, adaptive planning remains resource-intensive and difficult to implement routinely\cite{jafar2026,yu2025,hosny2018}. A typical workflow requires a complex, iterative decision-making process: clinicians must first strategize the optimal design for a specific fraction, then execute a suite of predictive tasks---including new volumetric imaging, deformable alignment, and complete re-segmentation of targets and OARs\cite{hosny2018,mccomas2023}. Crucially, this is often an iterative process\cite{keall2022}; if the initial predictions or dose distributions are found to be suboptimal, the team must loop back to redesign the plan, necessitating further manual adjustments. These steps require constant, high-level expert input from radiation oncologists, medical physicists, and dosimetrists, often taking hours---or even an entire day---depending on disease site and planning complexity (Fig.~\ref{fig:overview}d). This operational and cognitive burden has therefore limited widespread use of daily adaptive radiotherapy and remains a formidable barrier to routine clinical care.

Artificial intelligence has shown considerable promise in accelerating individual components of the adaptive radiotherapy workflow\cite{jafar2026,yu2025,hosny2018}, including image synthesis\cite{zhang2023breath,huijben2024}, image registration\cite{huang2022,nenoff2023}, auto-segmentation\cite{cao2017,poel2025,luo2025}, and dose prediction\cite{mcintosh2021,zhang2024dosediff,hu2021,jiao2023}. Yet most existing methods are designed for narrowly defined, isolated tasks. Consequently, they do not solve the main clinical problem: producing a complete, patient-specific adaptive treatment plan within the tight time constraints of treatment delivery. Another major limitation is that these models do not have the higher-level clinical intelligence required to organize these outputs into a practical treatment strategy. Because adaptive planning depends on the specific anatomical and clinical scenario, different patients and fractions may require different planning pathways. For example, synthetic CT may be required for dose recalculation in one fraction, or used mainly to support segmentation in another, or not needed at all when the original plan remains clinically acceptable. Isolated AI models therefore cannot autonomously decide whether their outputs are required in a given clinical scenario, how those outputs should be used, or how they should be assembled into a complete and executable treatment plan. As a result, even accurate single-task models have had limited translational impact, and their adoption has not yet enabled widespread online adaptive radiotherapy, where the full workflow must be completed rapidly, coherently, and with sufficient clinical reliability to support integrated decision-making.

\begin{figure}[H]
\centering
\includegraphics[width=0.95\textwidth]{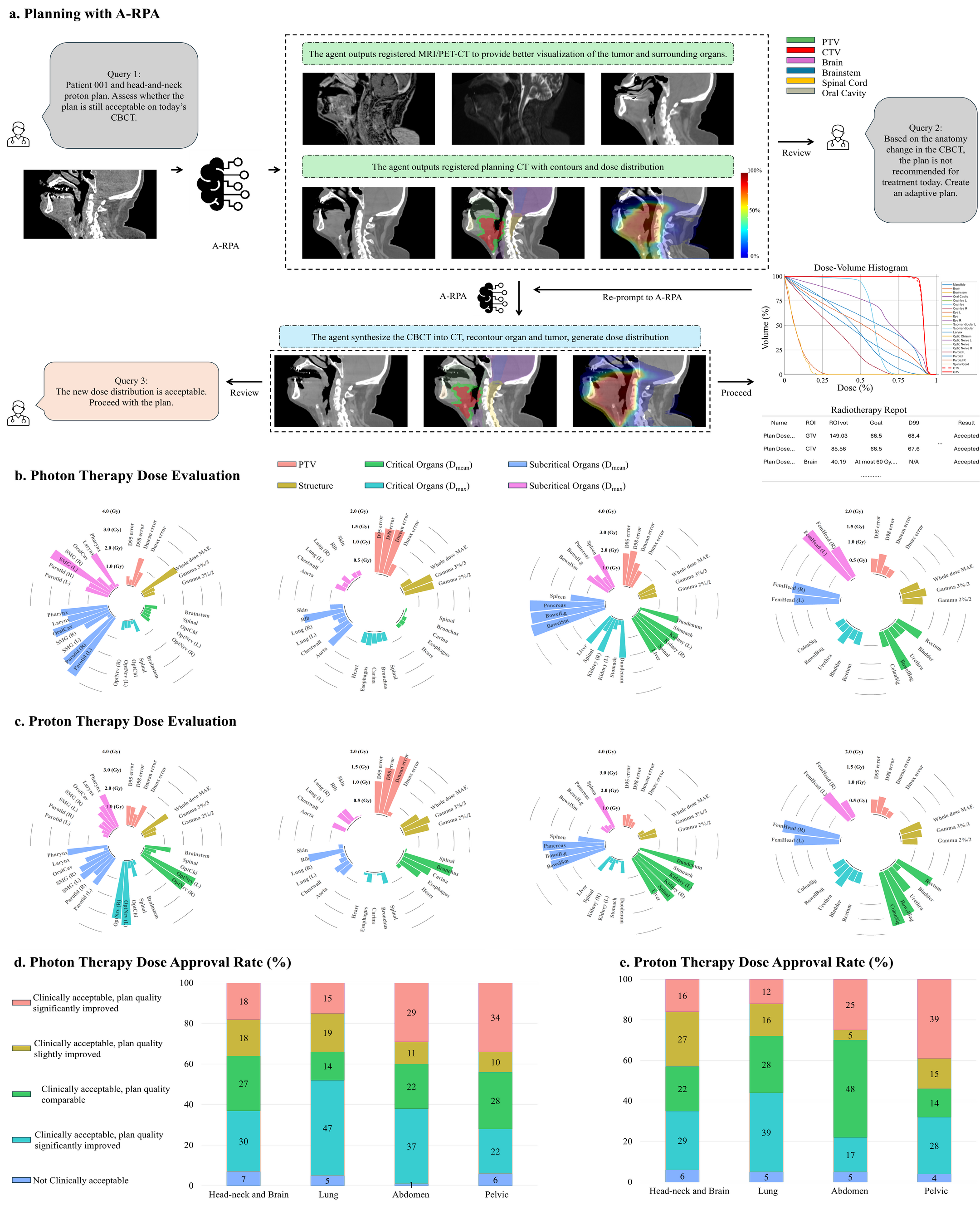}
\caption{\textbf{Representative clinician--agent interaction for head-and-neck adaptive radiotherapy planning on daily CBCT.} \textbf{a,} Representative head-and-neck case showing clinician-guided interaction with the agent on daily CBCT. The clinician first queries whether the original treatment plan remains acceptable on the anatomy observed on that day. In response, the agent registers prior multimodal images, including MRI or PET--CT, to the daily CBCT to improve visualization of the tumor and surrounding OARs, and propagates the planning CT, contours and dose distribution for plan assessment. When the anatomical change indicates that the original plan is no longer appropriate for treatment, the clinician issues a second query requesting adaptive replanning. The agent then synthesizes a CT from the daily CBCT, recontours targets and OARs, predicts a new dose distribution and returns an updated plan for evaluation. In this example, the replanned dose distribution is considered acceptable for treatment. \textbf{b,c,} Structure-wise dosimetric evaluation of agent-generated adaptive plans for photon radiotherapy \textbf{(b)} and proton radiotherapy \textbf{(c)}. Shown are planning target volume dose errors including $D_{95}$ and $D_{98}$ (in Gy), global dose-distribution error metrics, and organ-specific dose errors, including $D_{\mathrm{mean}}$ and $D_{\max}$, for critical and subcritical organs. \textbf{d,e,} Clinical acceptability of agent-generated adaptive plans for photon radiotherapy \textbf{(d)} and proton radiotherapy \textbf{(e)}, based on medical physicist review and summarized across acceptability categories.}
\label{fig:hn}
\end{figure}

Here, we present the One-for-All Adaptive Radiotherapy Planning Agent (A-RPA), an end-to-end Agentic AI system\cite{sapkota2025} that enables treatment-specific adaptive radiotherapy planning directly from the daily cone-beam CT (CBCT) acquired on the linear accelerator immediately before treatment. As shown in Fig.~\ref{fig:overview}, A-RPA combines two key components. The first is a unified One-for-All foundation model (Fig.~\ref{fig:overview}c) built from a pretrained foundation model (DINOv2)\cite{oquab2023}, designed to perform the major planning tasks required in adaptive radiotherapy within a single framework, including CT synthesis, multimodal image registration, tumor and OAR segmentation, and dose prediction. The second is an agent with a large language model (T5-XXL)\cite{raffel2020} (Fig.~\ref{fig:overview}b), designed to interpret the clinician's request, determine which model outputs are needed for the specific clinical scenario, and organize these outputs into an appropriate adaptive workflow. To do this, the agent integrates the daily CBCT with prior planning information, including planning CT or 4DCT, contours, prior dose, and available multimodal imaging such as MRI or PET-CT, and selects the most suitable strategy for that treatment fraction. Depending on the observed anatomy and clinical objective, the strategy may range from simple plan reuse, to deformable registration-based propagation of contours and dose, re-segmentation on registered images, or full replanning with CT synthesis, registration, segmentation, and dose prediction. In this way, A-RPA goes beyond isolated task prediction by both providing the full set of core radiotherapy planning functions and autonomously organizing them into a coherent, treatment-specific, and clinically reviewable adaptive plan.

\section*{Results}

\subsection*{A-RPA enables clinician-guided adaptive radiotherapy planning from daily CBCT}

We first assessed whether A-RPA could execute the sequence of decisions and predictions required for adaptive radiotherapy planning from a single daily CBCT and return outputs that were directly interpretable by clinicians during review. This evaluation was motivated by the central role of daily decision-making in adaptive radiotherapy, in which clinicians must determine not only whether the original plan remains acceptable based on the anatomy of the day, but also which adaptation pathway is required when anatomical change is observed. As shown in Fig.~\ref{fig:hn}a, A-RPA reproduced this clinical logic within a unified clinician--agent workflow. In response to an initial query on plan acceptability, the agent retrieved prior patient-specific planning information from hospital database, including the planning CT, historical contours, prior dose distribution and available multimodal images such as MRI or PET--CT, and registered these data to the daily CBCT to support assessment of target coverage and OAR alignment. At this stage, full synthetic CT generation, re-segmentation and dose recalculation were not required, because the immediate objective was to determine whether the existing plan could still be used safely. When the clinician identified anatomical change sufficiently to preclude direct plan reuse, a second query was issued to request generation of a new adaptive plan. In response, the agent initiated and executed a full replanning pathway, including CBCT-to-CT synthesis, target and OAR segmentation, and dose prediction, and returned an updated plan for evaluation. When the generated plan was considered acceptable, the synthesized CT, contours and dose were transferred to the clinical treatment planning system for final review and approval.

We next evaluated whether the adaptive plans generated by A-RPA achieved clinically meaningful dosimetric quality across treatment modalities. This analysis was designed to test whether the resulting plans preserved both target coverage and organ sparing in photon and proton radiotherapy. Because the generated dose distribution depended on the workflow selected by the clinician and the agent, the final output could correspond to a propagated registered plan, a recontoured plan on registered images, or a full synthetic-CT-based replanning result. Each generated distribution was compared with the clinically delivered reference dose distribution that was used to clinically treat the patient. As shown in Fig.~\ref{fig:hn}b,c, structure-wise dosimetric evaluation demonstrated consistently favorable adaptive plan quality in both modalities, including planning target volume (PTV) coverage errors, voxel-wise dose-distribution error metrics and organ-specific dose errors for critical and subcritical structures. For target evaluation, we reported $D_{95}$ and $D_{98}$ errors\cite{gao2023}, defined as the absolute differences between the generated and reference doses covering 95\% and 98\% of the PTV, respectively. To assess global agreement of the three-dimensional dose distribution, we used mean absolute error (MAE), which measures the average voxel-wise absolute dose difference, together with gamma analysis pass rates at 2\%/2~mm and 3\%/3~mm\cite{miften2018}, which quantify spatial and dosimetric agreement between generated and reference dose distributions under standard radiotherapy gamma tolerances. For OARs, we reported $D_{\mathrm{mean}}$ and $D_{\max}$ errors\cite{gao2023}, defined as the absolute differences in mean dose and maximum dose relative to the reference plan. In photon radiotherapy, PTV coverage remained close to the reference plan, with a cumulative all-fractions $D_{95}$ and $D_{98}$ errors ranging from 0.5--1.8~Gy and 0.7--2.0~Gy, respectively. These values represent whole-plan dose differences rather than per-fraction errors, and therefore correspond to substantially smaller deviations at individual fractions. The structure-level dose-distribution errors remained low, with MAE in the range of 0.5--2.0~Gy and high gamma agreement (plan-level dose-distribution agreement under the chosen gamma tolerances) under both 2\%/2~mm and 3\%/3~mm criteria. OAR evaluation further showed low $D_{\mathrm{mean}}$ and $D_{\max}$ errors across both critical and subcritical organs, with maximum errors of 2.5~Gy and 4.0~Gy, respectively. Similar findings were observed in proton radiotherapy, in which target coverage also remained stable, with $D_{95}$ and $D_{98}$ errors of 0.5--1.4~Gy and 0.7--1.6~Gy, respectively, and low MAE, high gamma agreement rates and low organ dose errors across evaluated structures. These findings indicate that A-RPA can generate adaptive plans with consistent dosimetric fidelity in both photon and proton settings.

We then assessed whether these outputs were clinically acceptable by expert review. This analysis was motivated by the fact that dosimetric agreement alone does not fully establish clinical utility, particularly in adaptive radiotherapy, where the practical question is whether a generated plan is usable in the clinic and whether it offers benefit relative to the original plan. To determine clinical usability, medical physicists assigned each case to one of five ordered outcome categories: rank 1, failure to satisfy at least one clinical dose constraint that was met by the original plan; rank 2, preservation of all originally satisfied constraints but with worse dose characteristics than the reference adaptive plan, meaning the plan is clinically acceptable but suboptimal relative to a full manual replan; rank 3, satisfaction of all relevant constraints with close agreement to the reference dose distribution, defined as meeting all clinical goals with dose differences $<$2~Gy for target metrics and $<$3~Gy for OAR metrics compared to the reference; rank 4, satisfaction of all constraints with close agreement and with at least one target or organ metric improved relative to the original plan, indicating dosimetric benefit beyond simply maintaining safety; and rank 5, overall visual and dosimetric assessment indicating that the generated plan was better than the original plan across multiple metrics. Under this definition, ranks 2--5 represent clinically acceptable plans that could be used for treatment, ranks 3--5 represent plans of quality comparable to or better than the reference manual planning, and ranks 4--5 indicate clear dosimetric benefit over the original plan. In photon radiotherapy, ranks 2--5 accounted for 93--99\% of evaluated cases across all anatomy sites, with 48--72\% falling into ranks 3--5 and 34--50\% reaching the improvement categories of ranks 4--5 (Fig.~\ref{fig:hn}d). In proton radiotherapy, ranks 2--5 accounted for 94--96\% of cases, with 56--78\% in ranks 3--5 and 28--54\% in ranks 4--5 (Fig.~\ref{fig:hn}e). Together, these results show that A-RPA can support both the upstream clinical decision of whether adaptive replanning is necessary and the downstream generation of adaptive plans that are frequently clinically acceptable and, in a substantial fraction of cases, superior to the original plan.

\begin{figure}[H]
\centering
\includegraphics[width=0.95\textwidth]{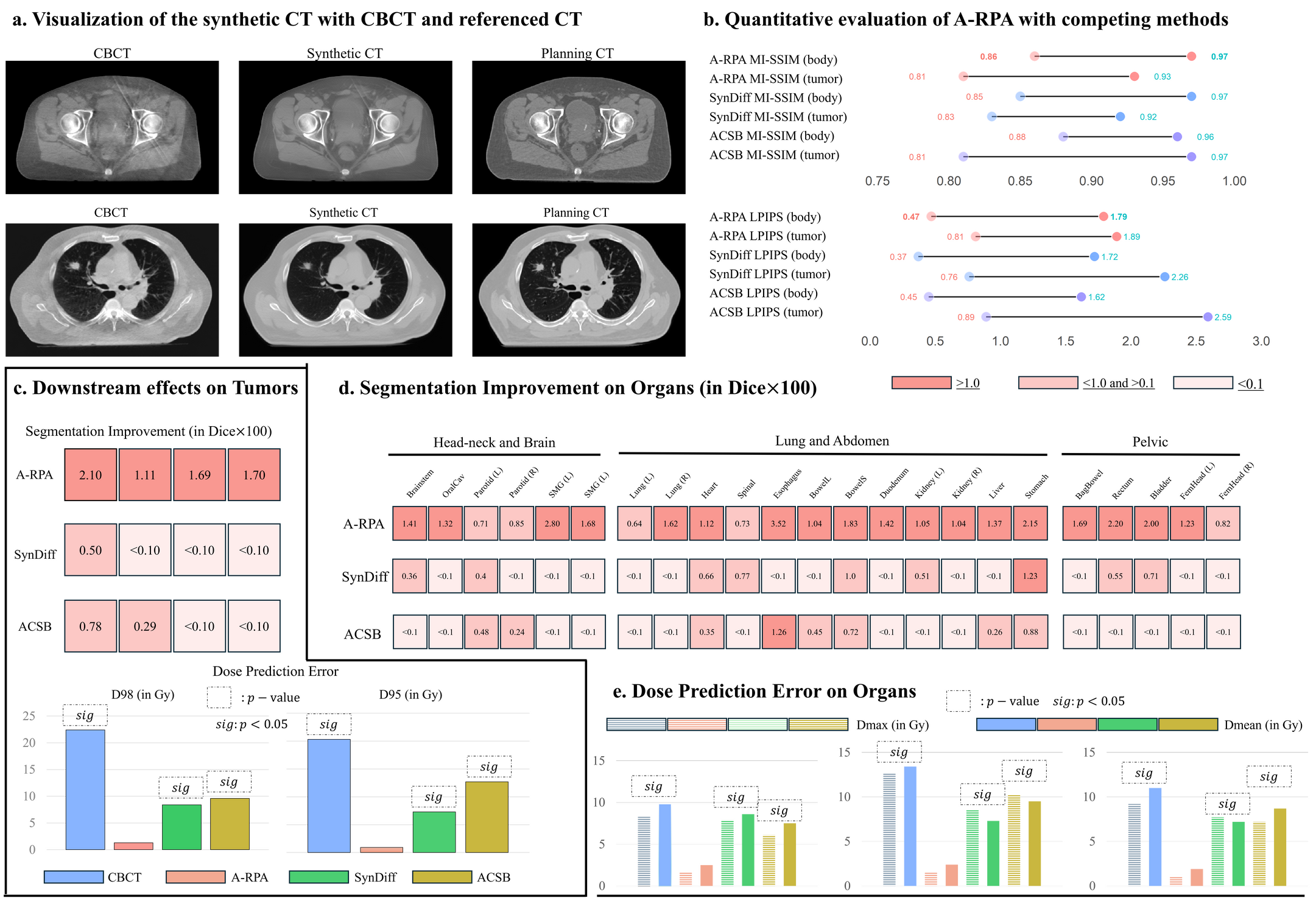}
\caption{\textbf{Synthetic CT quality and downstream benefit for adaptive radiotherapy planning.} \textbf{a,} Representative pelvic and lung cases showing the daily CBCT, the A-RPA synthetic CT and the corresponding reference planning CT. \textbf{b,} Quantitative image-quality comparison in body and tumor regions using MS-SSIM and LPIPS for A-RPA, and two competing unsupervised CBCT-to-CT synthesis methods including SynDiff and ACSB. \textbf{c,} Tumor-level analysis showing segmentation improvement relative to the original CBCT across different methods, with dose error ($D_{95}$ and $D_{98}$) compared among dose calculations performed on the CBCT, A-RPA and competing methods. Statistical significance between the baseline and competing methods was evaluated using a two-tailed Mann-Whitney U test, with p-values shown in the dotted boxes (``Sig'' indicates p-values $<$0.05 which is statistical significance). \textbf{d,} OARs-level segmentation improvement measured across a broad range of OARs, quantified as Dice improvement $\times$100. \textbf{e,} OARs-level improvement in dose error metrics, including $D_{\mathrm{mean}}$ and $D_{\max}$, compared across dose calculation on the original CBCT, A-RPA and competing methods. In \textbf{c--e}, dark red denotes segmentation improvement across various OARs greater than 1.0 with statistical significance, red denotes improvement between 0.1 and 1.0 with statistical significance, and a light color denotes improvement less than 0.1 or no significant improvement.}
\label{fig:sct}
\end{figure}

\subsection*{A-RPA's synthetic CT generation improves image fidelity and downstream planning utility}

We next examined whether synthetic CT generated from daily CBCT could provide sufficient image quality to support downstream adaptive planning in cases where anatomical change was substantial and registration-based adaptation was no longer adequate. In this workflow, synthetic CT was used when the anatomy of the day differed enough from the planning anatomy that direct registration and contour propagation could not reliably support replanning at the time of treatment. In representative pelvic and lung cases, A-RPA generated synthetic CTs with realistic anatomical appearance and clear preservation of clinically relevant structures visible on the daily CBCT (Fig.~\ref{fig:sct}a). We first quantified image-level similarity using multiscale structural similarity (MS-SSIM)\cite{wang2003}, which evaluates structural consistency and was computed between CBCT and synthetic CT to assess preservation of anatomical content. We also computed learned perceptual image patch similarity (LPIPS)\cite{zhang2018}, which measures perceptual discrepancy and was computed between reference CT and synthetic CT to assess visual realism. We compared A-RPA with two competing unsupervised CBCT-to-CT synthesis methods, SynDiff\cite{ozbey2023} and adversarial consistency-based synthesis (ACSB)\cite{shi2025}. Across body and tumor regions, these visual metrics were broadly similar among A-RPA, SynDiff and ACSB, with values lying in a relatively narrow range (Fig.~\ref{fig:sct}b). These results indicate that all three methods produced visually plausible synthetic CTs, and that conventional image-level metrics alone did not fully distinguish their downstream clinical value.

The more substantial differences emerged in segmentation and dose evaluation. At the tumor level, A-RPA consistently produced substantially larger segmentation gains than competing methods across all evaluated disease sites, with Dice similarity coefficient (Dice) improvements of 2.10, 1.11, 1.69 and 1.70 (all expressed as Dice $\times$100), compared with negligible gains from competing methods: only 0.50, $<$0.10, $<$0.10 and $<$0.10 for SynDiff, and 0.78, 0.29, $<$0.10 and $<$0.10 for ACSB (Fig.~\ref{fig:sct}c). Dice quantifies spatial overlap between predicted and reference contours, with higher values indicating more accurate segmentation. The same pattern was observed for dose assessment. Direct dose calculation on the original unprocessed CBCT resulted in prohibitively large tumor dose errors, with mean absolute errors of approximately 22.4~Gy and 20.5~Gy for the two reported metrics. A-RPA reduced these errors dramatically to approximately 1.2~Gy and 0.8~Gy, whereas SynDiff reduced them only to approximately 8.5~Gy and 7.2~Gy, and ACSB to approximately 9.7~Gy and 12.3~Gy, where they are all still within a range that would be clinically unacceptable for treatment delivery (Fig.~\ref{fig:sct}c). Crucially, statistical analysis using two-tailed Mann-Whitney U test confirmed that the performance gains achieved by A-RPA in both tumor segmentation and dose reduction were highly significant compared with all competing methods across every evaluation metric ($p<0.05$). Thus, although competing synthesis methods improved visual image quality as measured by conventional metrics (Fig.~\ref{fig:sct}b), A-RPA yielded substantially greater benefit for tumor-relevant segmentation and dose recovery---a direct consequence of its joint optimization within the unified framework, which ensures that intermediate representations remain informative for downstream clinical tasks.

This advantage became even more pronounced in OARs-level analysis. Across a broad range of OARs, A-RPA produced consistently and substantially larger segmentation improvements than competing methods, with Dice gains ranging from 0.64 to 3.52 points and multiple structures showing gains above 2.0, including values of 2.80, 3.52, 2.15, 2.20 and 2.00 (Fig.~\ref{fig:sct}d). By contrast, SynDiff showed only modest gains, mostly below 1.23, and ACSB showed gains mostly below 1.26, with many OARs showing minimal or negligible change ($<$0.10). OARs-level dose analysis showed a similar trend. When dose was evaluated directly on CBCT, OARs dose errors remained high, generally ranging from 8--13~Gy across the evaluated structure, which were well outside clinically acceptable tolerances. A-RPA reduced these errors dramatically to approximately 0.6--2.5~Gy (Fig.~\ref{fig:sct}e). SynDiff and ACSB provided only partial correction, with residual errors still generally in the 6--10~Gy range, which are insufficient for reliable clinical use in an adaptive setting. Taken together, these findings show that the main advantage of A-RPA is not merely improved synthetic CT appearance, but stronger downstream clinical utility. By jointly optimizing synthesis, registration, segmentation and dose modeling within a unified One-for-All framework with shared gradient propagation during training, A-RPA converts daily CBCT into a representation that supports substantially more accurate segmentation and dose evaluation than competing unsupervised synthesis approaches, directly validating the central hypothesis that end-to-end optimization across the entire adaptive planning cascade yields greater clinical value than optimizing individual components in isolation.

\subsection*{A-RPA's multimodal registration improves anatomical alignment and downstream segmentation}

\begin{figure}[H]
\centering
\includegraphics[width=0.95\textwidth]{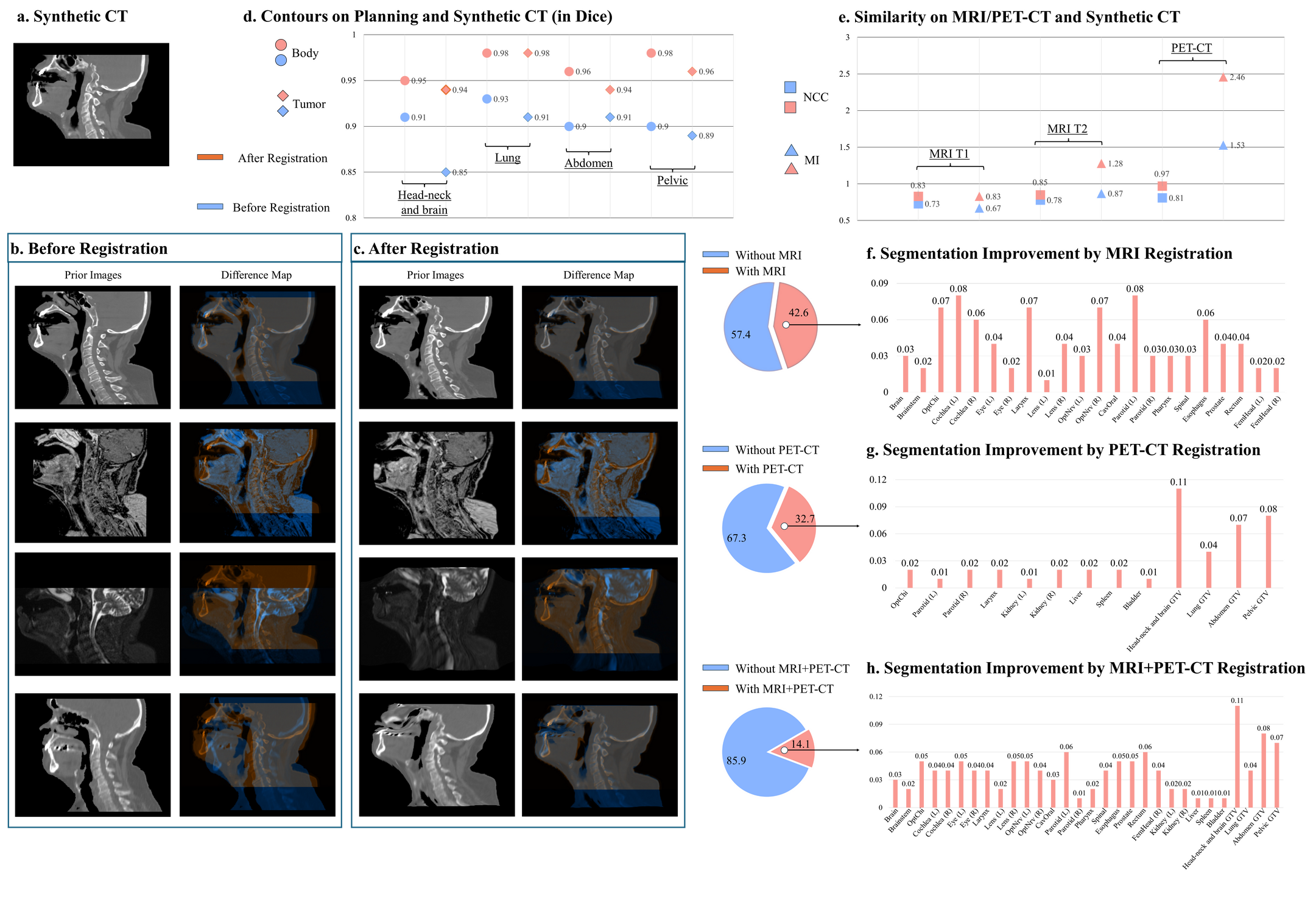}
\caption{\textbf{Multimodal registration quality and downstream benefit for adaptive radiotherapy planning.} \textbf{a,} Representative target anatomy and quantitative summary of registration performance. Dice similarity coefficient, which measures spatial overlap between contours, is shown before and after registration for organ and tumor structures, using contours from the planning CT and the physician-approved registered CT as reference. Normalized cross-correlation and mutual information, which quantify image similarity across modalities, are reported for MRI and PET--CT, for which structure annotations were not available. \textbf{b,} Qualitative examples of multimodal alignment before registration, shown for planning CT, MRI and PET--CT with contour overlays in the target space. \textbf{c,} Qualitative examples of multimodal alignment after registration, shown for planning CT, MRI and PET--CT with contour overlays in the target space. \textbf{d,} Case-level contour agreement before and after registration, evaluated by Dice similarity coefficient against the physician-approved registered CT. \textbf{e,} Registration quality for MRI and PET--CT evaluated by NCC and MI. \textbf{f--h,} OARs-level downstream segmentation improvement after registration using MRI only \textbf{(f)}, PET--CT only \textbf{(g)} and MRI plus PET--CT \textbf{(h)}; the accompanying pie charts show the proportion of evaluated cases with and without the corresponding modality configuration.}
\label{fig:reg}
\end{figure}

Because target extent and soft-tissue anatomy are often more clearly defined on MRI or PET--CT than on CBCT alone, we next examined whether incorporating these complementary imaging modalities through deformable registration could improve anatomical alignment in the adaptive setting, and whether such improvements translated into downstream segmentation benefits. In this workflow, registration was primarily used to wrap and integrate complementary pre-treatment imaging into the synthetic CT of the day, thereby providing enhanced soft-tissue context that is not available from CBCT alone. Its role was to support both clinical visualization for physician review and downstream segmentation for automated contour adaptation. Representative examples showed that registration successfully brought the planning CT, MRI and PET-CT into closer correspondence with the anatomy of the day, with visibly improved alignment of contour boundaries and multimodal image features in the target space (Fig.~\ref{fig:reg}b,c). This improved alignment is clinically significant because it enables more accurate target definition and OAR sparing during adaptive replanning.

We first quantified registration quality using the Dice similarity coefficient, which measures spatial overlap between transformed and reference contours, with higher values indicating better geometric agreement. Relative to the pre-registration state, registration increased body-contour Dice from 0.90--0.92 to 0.93--0.98 across the four anatomical groups, and tumor-contour Dice from 0.85--0.90 to 0.90--0.95 (Fig.~\ref{fig:reg}a,d). The largest gains were observed in the thoracic (lung) and pelvic settings, where anatomical deformation is often most pronounced: body Dice reached approximately 0.98 after registration in both sites and tumor Dice increased to approximately 0.95, representing near-complete geometric alignment with reference contours. These gains are clinically meaningful, as residual misalignment of even a few millimeters can substantially affect dose calculation accuracy in both photon and proton therapy.

For MRI and PET-CT, which did not contain contour annotations for all cases, we additionally evaluated registration quality using image-based similarity metrics. We used normalized cross-correlation (NCC), which reflects agreement in local intensity patterns, and mutual information (MI), which captures statistical dependence between images and is widely used for cross-modality alignment. Both metrics improved after registration across all evaluated modalities (Fig.~\ref{fig:reg}a,e). For MRI T1, NCC increased from 0.73 to 0.83, and MI increased from 0.67 to 0.83. For MRI T2, NCC increased from 0.79 to 0.86, and MI increased from 0.87 to 1.26. For PET-CT, NCC increased from 0.81 to 0.97, and MI increased from 1.53 to 2.46. These results indicate that the registration framework improved not only contour-level geometric agreement, but also broader image-level correspondence across modalities.

These alignment gains directly translated into downstream benefits in segmentation. For downstream analysis, each modality-specific comparison was performed in a paired manner within the subset of patients in whom that modality was available; for example, MRI-guided benefit was evaluated by comparing segmentation performance with and without the inclusion of registered MRI in the same set of MRI-available cases, with analogous comparisons performed for PET--CT and for cases where both MRI and PET-CT were available. Under this rigorous paired setting, MRI-guided registration improved segmentation across multiple structures, with raw Dice score gains ranging from 0.01 to 0.09 (Fig.~\ref{fig:reg}f). While these numerical gains may appear modest, they are clinically meaningful in soft-tissue structures where baseline CBCT contrast is inherently limited. PET-CT-guided registration also improved segmentation, with gains ranging from 0.01 to 0.11, with larger effects in selected target-related structures, reflecting the functional and metabolic information captured by PET that can refine tumor boundaries (Fig.~\ref{fig:reg}g). The broadest benefit was observed when MRI and PET--CT were used together, with improvement distributed across a substantially wider set of OARs and targets, suggesting synergistic value from integrating complementary anatomical (MRI) and functional (PET) information (Fig.~\ref{fig:reg}h). The accompanying modality distributions indicate that these analyses reflect real-world imaging availability, with MRI available in 42.6\% of evaluated cases, PET--CT in 32.7\%, and both MRI and PET--CT in 14.1\%. Together, these findings show that multimodal registration can successfully project complementary anatomical and functional information into the anatomy of the day and consistently improve downstream segmentation performance beyond CBCT-only assessment, thereby addressing one of the fundamental limitations of CBCT-guided adaptive radiotherapy: its inherently poor soft-tissue contrast.

\begin{figure}[H]
\centering
\includegraphics[width=0.95\textwidth]{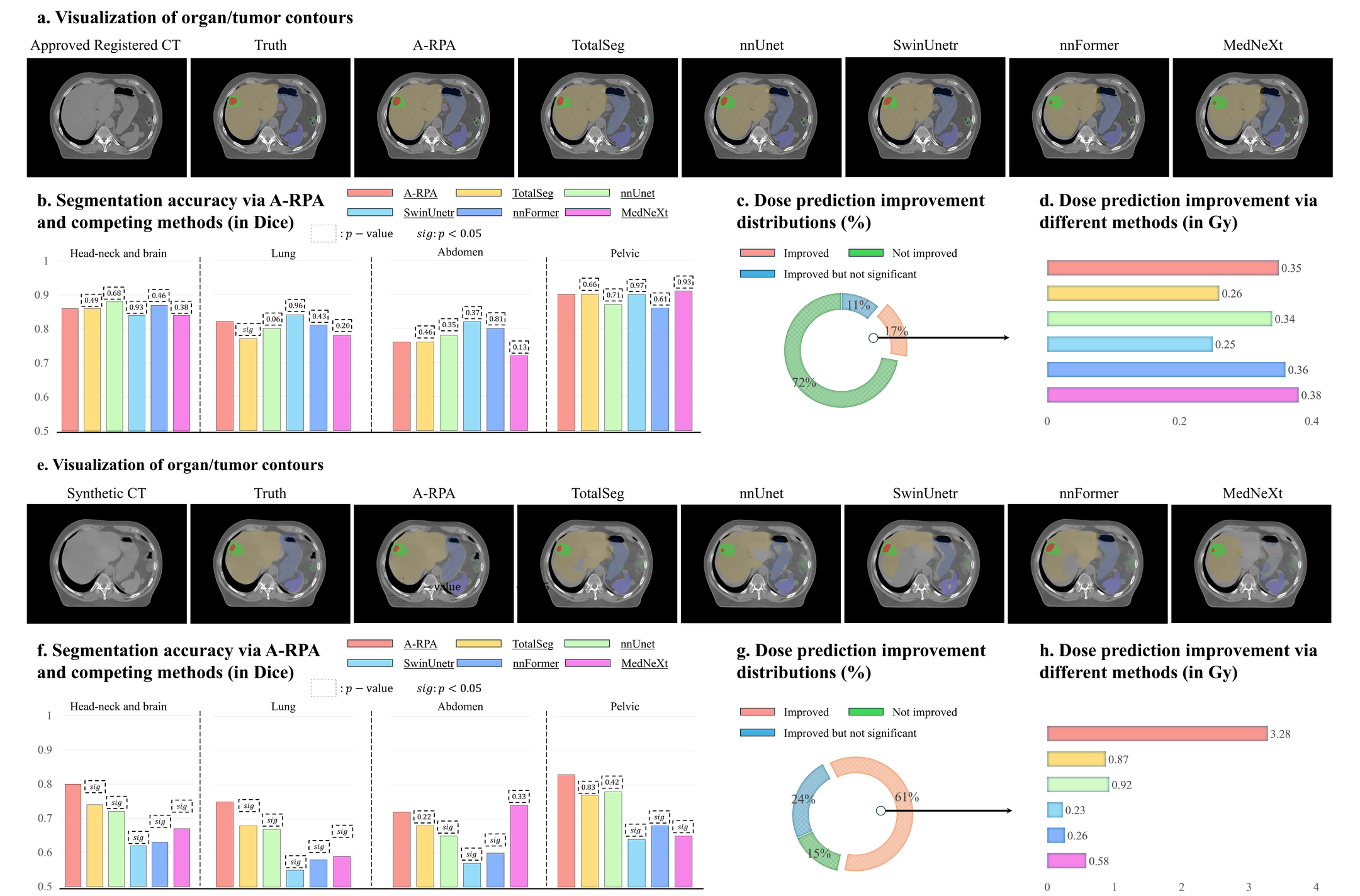}
\caption{\textbf{Segmentation quality and downstream benefit for adaptive radiotherapy planning.} \textbf{a,} Representative visualizations of A-RPA and competing segmentation methods on registered CT. \textbf{b,} Segmentation performance on registered CT across head-and-neck, thoracic, abdominal and pelvic radiotherapy, reported by Dice similarity coefficient. Statistical significance between the A-RPA and competing methods was evaluated using a two-tailed Mann-Whitney U test, with p-values shown in the dotted boxes (``Sig'' indicates p-values $<$0.05 which is statistical significance). \textbf{c,} Proportion of cases in which downstream dose evaluation changed when re-segmentation was used instead of direct contour propagation from the registered planning CT. \textbf{d,} Magnitude of dose improvement among affected cases in \textbf{c}, stratified by method. \textbf{e,} Representative visualizations of A-RPA and competing segmentation methods on synthetic CT. \textbf{f,} Segmentation performance on synthetic CT across head-and-neck, lung, abdominal and pelvic radiotherapy, reported by Dice similarity coefficient. Statistical significance between the A-RPA and competing methods was evaluated with p-values shown in the dotted boxes (``Sig'' indicates statistical significance). \textbf{g,} Proportion of cases in which downstream dose evaluation changed when synthetic-CT-based re-segmentation was used instead of direct use of the contours from the registered planning CT alone. \textbf{h,} Magnitude of dose improvement among affected cases in \textbf{g}, stratified by method. In \textbf{c} and \textbf{g}, red denotes cases with improvement and statistical significance, blue denotes cases with improvement without statistical significance, and the remaining category denotes cases without improvement.}
\label{fig:seg}
\end{figure}

\subsection*{A-RPA's segmentation improves OARs/tumor contour accuracy and downstream dose prediction}

We evaluated segmentation on both registered CT and synthetic CT because they capture the two principal re-segmentation scenarios in adaptive radiotherapy: cases in which the aligned planning CT (after deformable registration) remains visually acceptable but propagated contours are inadequate and require re-segmentation, and cases in which anatomical change necessitates synthetic-CT-based re-segmentation for full downstream replanning. This distinction is clinically important because the former scenario represents a lighter adaptation (contour touch-up on a reliable anatomical substrate), whereas the latter represents a complete replanning pathway where the underlying image itself is generated and therefore potentially less reliable. We first evaluated the segmentation component of A-RPA against competing networks, including TotalSegmentator\cite{wasserthal2023}, 3D nnU-Net\cite{isensee2024}, Shifted Window U-Net with Transformer Encoder (SwinUNETR)\cite{hatamizadeh2021}, interleaved Transformer for volumetric segmentation (nnFormer)\cite{zhou2021}, and medical ConvNeXt-based segmentation network (MedNeXt)\cite{roy2023} on registered CT. Representative examples across treatment sites showed that A-RPA produced anatomically accurate contours on registered CT (Fig.~\ref{fig:seg}a). We quantified segmentation accuracy using the Dice. Across head-and-neck, lung, abdominal and pelvic radiotherapy, segmentation performance on registered CT remained consistently high for all methods, with Dice values ranging from 0.82 to 0.94 across structures, with no method consistently outperforming others by more than 0.02 (Fig.~\ref{fig:seg}b). Statistical evaluation using two-tailed Mann-Whitney U test confirmed that A-RPA did not show any significant difference compared with the competing methods across these configurations ($p>0.05$). These results indicate that once registered CT anatomy has been aligned to the daily geometry with high fidelity, re-segmentation becomes a relatively well-posed problem for which differences among segmentation models are limited.

We then examined whether direct contour propagation from the planning CT was sufficient for adaptive dose assessment, or whether explicit re-segmentation still provided added value. As expected (Fig.~\ref{fig:seg}c,d), because the input remained a registered real CT, only a modest subset of cases benefited from re-segmentation. Nevertheless, a subset of cases still showed altered downstream dose evaluation when re-segmentation was used instead of directly reusing propagated contours (Fig.~\ref{fig:seg}c), indicating that residual misregistration can remain clinically relevant. In this setting, A-RPA yielded the largest proportion of cases with no improvement, at approximately 72\%. About 17\% showed significant improvement ($>$0.1~Gy), and 11\% showed smaller, non-significant improvement. Among affected cases, the magnitude of improvement was also limited, with values around 0.35--0.40~Gy across methods (Fig.~\ref{fig:seg}d). Thus, on registered CT, re-segmentation can still be beneficial, but the gain remains moderate because the underlying image is already a transformed real CT.

We next evaluated segmentation on synthetic CT, which is the image representation used when full adaptive replanning is required. This setting is fundamentally more challenging: although synthetic CT visually resembles CT, it is still a generated representation rather than a real one, and accurate segmentation therefore depends not only on segmentation ability, but also on how well the synthesized anatomy preserves features required for downstream delineation. Representative examples showed that A-RPA produced contours with closer anatomical agreement to the reference than competing methods on synthetic CT (Fig.~\ref{fig:seg}e). Quantitative evaluation again used Dice similarity coefficient as the primary metric (Fig.~\ref{fig:seg}f). In contrast to the registered-CT setting, where all methods performed similarly, A-RPA retained strong performance on synthetic CT across head-and-neck, lung, abdominal and pelvic radiotherapy, with Dice values of approximately 0.71, 0.74, 0.72 and 0.82, respectively, whereas competing methods showed substantially larger degradation in several treatment sites. Statistical analysis confirmed that the robust performance maintained by A-RPA was significantly superior to all competing methods across the evaluated sites ($p<0.05$). These findings indicate that synthetic CT segmentation is not simply a standard CT segmentation problem, and that only a method jointly optimized with the synthesis process can maintain robust downstream contour accuracy.

We further assessed whether synthetic-CT-based re-segmentation affected downstream dose evaluation. As in the registered-CT setting, dose assessment changed in a meaningful proportion of cases when re-segmentation was performed instead of relying on transferred contours alone (Fig.~\ref{fig:seg}g). However, the magnitude of benefit was much larger on synthetic CT. A-RPA showed statistically significant improvement in approximately 61\% of cases, with an additional 15\% showing smaller non-significant improvement, whereas 24\% showed no improvement. Among the affected cases, A-RPA produced a markedly larger dose improvement, reaching approximately 3.63~Gy, whereas competing methods remained much lower, at approximately 0.67~Gy, 0.92~Gy, 0.23~Gy, 0.36~Gy, and 0.58~Gy (Fig.~\ref{fig:seg}h). Together, these results show that segmentation on registered CT is reliable, whereas segmentation on synthetic CT more clearly reveals the advantage of A-RPA. By coupling synthesis and segmentation within a unified One-for-All framework, A-RPA produces synthetic CT representations that are not only visually realistic, but also substantially more suitable for downstream segmentation and dose evaluation.

\section*{Discussion}

The value of a unified foundation model for adaptive radiotherapy lies not only in its generalizability across heterogeneous clinical settings, but also in its potential to make daily adaptation practically achievable within the time constraints of treatment delivery. In current clinical practice, adaptive replanning often requires extensive coordination among physicians, physicists and dosimetrists, and may delay treatment by several hours or even longer, making routine fraction-by-fraction adaptation difficult to sustain. In this study, A-RPA addressed this barrier by integrating synthesis, registration, segmentation and dose prediction into a single agent-guided workflow that operated in approximately 2~min, which is compatible with online adaptive workflows. Across photon and proton radiotherapy, the resulting adaptive plans maintained low target dose errors, with $D_{95}$ and $D_{98}$ errors generally within 0.5--2.0~Gy of the clinically delivered reference plans, supporting the feasibility of rapid yet clinically meaningful adaptive planning directly from the daily CBCT without compromising dosimetric accuracy.

A key insight from this work is that an agentic framework is necessary because adaptive radiotherapy does not follow a single fixed workflow. Depending on the anatomy of the day, clinicians may choose simple plan reassessment, multimodal registration for improved visualization, re-segmentation on registered CT when propagated contours are inadequate, or full synthetic-CT-based replanning when anatomical change is too large for registration-based adaptation alone. Selecting among these pathways is therefore itself a clinical reasoning problem. In this study, the agent was designed not merely to execute models, but to interpret clinician intent, retrieve relevant prior information and trigger the appropriate planning pathway for the specific fraction. This design yielded outputs that were not only dosimetrically accurate, but also frequently clinically acceptable, with 93--99\% of photon cases and 94--96\% of proton cases preserving clinically acceptable planning constraints. Thus, the principal value of the agent is not simply automation, but orchestration of branching clinical workflows in a form that remains aligned with how adaptive planning is actually performed in practice, a distinction that positions A-RPA as a clinical decision support system rather than merely a pipeline of automated tools.

A central finding of this work is that a unified, One-for-All architecture can leverage shared anatomical representation learning while preserving task-specific specialization, and that this coupling provides clinical advantages beyond operational convenience alone. The DINOv2-based pathway provided a common anatomical encoder for multimodal registration and contour adaptation, whereas the diffusion-based pathway supported generative synthesis of synthetic CT and three-dimensional dose. More importantly, the unified design allowed gradients and task information to propagate across components during training, so that upstream representations remained informative for downstream planning tasks. This appears to be particularly important for synthetic CT. Although competing synthesis methods produced visually plausible images, only jointly optimized A-RPA consistently preserved downstream utility. For example, tumor dose errors were reduced from approximately 22~Gy on CBCT to approximately 1~Gy with A-RPA, compared with 7--12~Gy for competing synthesis methods. Likewise, on synthetic CT, A-RPA produced markedly larger downstream dose improvement (3.63~Gy) than competing methods (0.23--0.92~Gy). These findings suggest that separately optimized modules may generate outputs that appear realistic in isolation, yet remain suboptimal for downstream segmentation and dose evaluation, whereas joint optimization enables intermediate representations to remain clinically useful for the next planning step.

These findings suggest that A-RPA presents a practical pathway for routinely implementing online adaptive radiotherapy by enabling plan assessment and replanning directly from daily CBCT, reducing reliance on repeat CT simulation and extensive manual re-segmentation. The framework can also bring MRI or PET--CT information into the treatment-day setting when additional anatomical or functional context is needed. Further, A-RPA preserves clinical oversight through physician-guided review of intermediate outputs and final approval. More broadly, the use of complete treatment-course information for joint training allowed synthesis, registration, segmentation and dose prediction to be optimized within one adaptive workflow, which likely contributed to the observed downstream gains in segmentation and dose evaluation.

The study also highlights the promise of agentic artificial intelligence in radiotherapy. Most prior work in this area has focused on accelerating one planning component, such as auto-segmentation or dose prediction. Although valuable, these task-specific systems do not directly address the core challenge of real-time adaptive planning, in which the full chain of operations must be selected, executed and interpreted under clinical time pressure. By contrast, A-RPA frames adaptive radiotherapy as a reasoning and orchestration problem: the system must determine whether simple plan reassessment is sufficient, whether registration-guided adaptation is needed, whether re-segmentation on registered CT is required, or whether full replanning should be triggered. This design more closely reflects real clinical decision-making and suggests a path toward future radiotherapy systems that combine foundation models with structured reasoning and human-in-the-loop control.

Despite these strengths, several limitations merit discussion. First, although the framework generalized across multiple anatomical sites and treatment modalities, all joint post-training and evaluation were performed using CBCT scans of clinically acceptable quality. Performance deteriorates substantially when CBCT quality is poor, as severe artifacts or low image fidelity can degrade synthetic CT generation and propagate errors through downstream registration, segmentation and dose prediction. In addition, when the CBCT has a limited field of view, such as a half-fan acquisition, even a visually high-quality synthetic CT may remain insufficient for dose calculation because important anatomy outside the scanned region is not captured. Furthermore, the CBCT data in this cohort were acquired under relatively consistent institutional practice and were generally of sufficient quality for adaptive review. Performance may therefore differ in external settings with other scanner vendors, reconstruction pipelines, segmentation practices or more challenging image characteristics, including more severe artifact and lower soft-tissue contrast on CBCT\cite{jiang2022,xu2023,wilson2025}. Broader multi-institutional validation will be necessary to establish robustness across these sources of variation. Second, dose prediction remains an approximation of the clinical planning process. Although the generated dose distributions were clinically coherent, they do not fully replace inverse optimization\cite{yu2025,buchanan2023,sun2022} within a treatment-planning system and remain dependent on implicit learning of beam geometry, modality-specific practice and physician planning preferences. Incorporating more explicit physics-based constraints, machine delivery parameters or optimization-informed supervision may further improve robustness and clinical interpretability.

There are also architectural questions that remain open. We used a shared foundation framework with separate diffusion and anatomical pathways, but it is not yet clear whether performance is saturated with the present model scale, dataset diversity or conditioning strategy. As in other domains, further scaling may continue to improve performance\cite{vorontsov2024,gao2023synth,tham2025,ding2025,thapa2026}. In adaptive radiotherapy, however, the most effective axis of scaling remains uncertain. Increased anatomical diversity, larger longitudinal cohorts, broader variation in CBCT quality and richer multimodal supervision may all be important, particularly for rare treatment scenarios or highly variable tumor presentations. In addition, the current study used a cascaded supervision design to reflect the dependency structure of clinical planning. Future work should investigate whether tighter end-to-end coupling (e.g., parallel training\cite{fan2024,gupta2022}), uncertainty-aware optimization\cite{ai2024,liu2024} or reinforcement-based planning refinement\cite{tang2024,hong2025,uehara2024,fan2023,hiranaka2024,li2024} can further improve downstream clinical utility.

Looking forward, the A-RPA framework provides a foundation for next-generation adaptive radiotherapy systems that are not only faster, but also more context-aware and clinically interactive. Future extensions could incorporate temporal learning across fractions to model anatomical trends and anticipate future changes, uncertainty estimation for clinician-facing decision support and tighter integration with treatment-planning systems for closed-loop plan optimization. More broadly, the present study suggests that adaptive radiotherapy may benefit from the same shift seen in other areas of artificial intelligence: from narrow task-specific models toward general-purpose foundation systems that can support multiple downstream objectives within a unified framework. As radiotherapy increasingly moves toward individualized, daily adaptive treatment, such systems may help reduce manual workload, improve planning consistency and expand access to high-quality adaptive care across diverse disease sites and treatment settings.

\section*{Ethical approval}

The use of internal clinical data was approved by the Institutional Review Board of Emory University (Decatur, USA). All data were de-identified before analysis, and no patient was prospectively enrolled. The generated treatment plans from the A-RPA were used for retrospective research only and not for clinical delivery. Informed consent was waived in accordance with institutional and regulatory policies.

\section*{Data Availability}

Some of the pretraining datasets were retrospectively analyzed from publicly available databases, which can be accessed at the following addresses: CMB-PCA (\url{https://www.cancerimagingarchive.net/collection/cmb-pca/}), TotalSegmentator (\url{https://zenodo.org/records/14710732}), BIMCV-Prostate (\url{https://zenodo.org/records/13254318}), PROSTATE-MRI (\url{https://www.cancerimagingarchive.net/collection/prostate-mri/}), Prostate-3T (\url{https://www.cancerimagingarchive.net/collection/prostate-3t/}), PROSTATE-DIAGNOSIS (\url{https://www.cancerimagingarchive.net/collection/prostate-diagnosis/}), QIN-PROSTATE-Repeatability (\url{https://www.cancerimagingarchive.net/collection/qin-prostate-repeatability/}), fastMRI (\url{https://fastmri.med.nyu.edu/}), PROSTATEx (\url{https://www.cancerimagingarchive.net/collection/prostatex/}), and MSD (\url{http://medicaldecathlon.com/}).

\section*{Declaration of Competing Interest}

The authors declare that they have no known competing financial interests or personal relationships that could have appeared to influence the work reported in this paper.

\section*{Acknowledgement}

This research is supported in part by the National Institutes of Health under Award Number R01CA272991, R01DE033512, R56EB033332, R01EB032680, and P30CA008748.

\section*{Methods}

\subsection*{Datasets}

Data used in this study were organized into two main groups: public datasets for foundation model development and an internal clinical dataset for post-training of the unified A-RPA model. The public datasets included a foundation pre-training cohort and task-specific fine-tuning cohorts for unsupervised CBCT-to-CT synthesis, multimodal registration, OARs segmentation, and dose prediction. These datasets were considered together because some foundation pre-training datasets also provided task-specific annotations for downstream supervised adaptation. A complete listing of all public datasets, including their sources, modalities, and primary uses, is provided in the Data Availability section.

The internal dataset comprised complete, longitudinal radiotherapy treatment data from patients treated at Emory University Hospital and was used for joint post-training of the A-RPA model across all four core tasks simultaneously. The internal clinical cohort comprised 814 treatment courses from Emory University Hospital, spanning five major anatomical sites and both photon and proton radiotherapy workflows. This cohort included head-and-neck ($n=265$), brain ($n=62$), lung ($n=97$), abdomen ($n=182$) and pelvic/prostate ($n=208$) cases, with 460 photon and 354 proton treatment courses. Available imaging across the cohort included 814 planning CT scans, 17,398 CBCT scans, 284 T1-weighted MRI scans, 311 T2-weighted MRI scans and 294 PET-CT scans. The site-specific distribution comprised 160 photon and 105 proton head-and-neck cases, 36 photon and 26 proton brain cases, 71 photon and 26 proton lung cases, 97 photon and 85 proton abdominal cases, and 116 photon and 92 proton pelvic/prostate cases.

For model development, 70\% of the public datasets were used for training, and the remaining cases were assigned to validation and testing according to dataset-specific design. Likewise, 70\% of the internal clinical cohort was used for training, and the remainder was used for validation and evaluation. All CT, MRI and PET-CT images were resampled to $1 \times 1 \times 2$~mm resolution and padded to $512 \times 512 \times 256$. CT and CBCT intensities were clipped to the range of $-1024$ to 3,000~HU and normalized to $-1$ to 1. MRI and PET images were normalized to the range of $-1$ to 1.

\subsection*{Foundation pre-training and fine-tuning datasets}

As shown in Fig.~\ref{fig:overview}a, public datasets were grouped by anatomical region and modality coverage. Whole-body datasets included CT-ORG and TotalSegmentator, contributing 1,368 CT scans and supporting OARs segmentation. Head-and-neck and brain datasets included HaN-Seg, Head-Neck Cetuximab, SegRap2023, OpenKBP, RADCURE, HNSCC, ACSwinNet, SynthRAD2025, GDP-HMM and the brain CBCT--CT subset of SynthRAD2023, comprising 9,391 scans or studies (8,875 CT, 460 CBCT and 56 MRI) and supporting segmentation, multimodal registration, image synthesis and dose prediction. Thoracic datasets included NLST, DIR-Lab Lung 4DCT, 4D-Lung, Learn2Reg ThoraxCBCT, SynthRAD2025 and GDP-HMM, comprising 19,849 scans or studies (19,743 CT and 106 CBCT) and supporting registration, synthesis and dose prediction. Abdominal datasets included WORD, Pancreas-CT, BTCV, CHAOS, LiTS, KiTS, AbdomenCT-1K, AMOS22, MSD and Pancreatic-CT-CBCT-SEG, comprising 4,010 scans (3,790 CT, 80 CBCT and 140 MRI) and supporting segmentation and multimodal registration. Prostate and pelvic datasets included BIMCV-Prostate, PROSTATE-MRI, Prostate-3T, PROSTATE-DIAGNOSIS, QIN-PROSTATE-Repeatability, PROSTATEx, CMB-PCA, Learn2Reg cervix and the pelvic CBCT--CT subset of SynthRAD2023, comprising 5,461 scans or studies (320 CT, 270 CBCT and 4,871 MRI) and supporting segmentation and registration. Additional auxiliary datasets, including IXI, OASIS, ACDC, fastMRI, AutoPET and the MRI-to-CT subset of SynthRAD2023, contributed 7,477 scans or studies (6,463 MRI, 540 CT and 1,014 PET-CT) and were used only for foundation pre-training to increase modality diversity and anatomical coverage.

Across these public datasets, scans with paired or unpaired cross-modality structure were further used for downstream tasks (Fig.~\ref{fig:overview}b), including unsupervised CBCT-to-CT synthesis, multimodal registration, OARs segmentation and dose prediction, with each dataset contributing to one or more tasks based on the availability of corresponding annotations and metadata. The diversity and scale of these public datasets provided the foundation for learning generalizable representations across anatomical sites, imaging modalities, and treatment planning tasks, prior to task-specific fine-tuning and joint post-training on the internal clinical cohort.

\subsection*{Overview of One-for-All Adaptive Radiotherapy Planning Agent}

We developed an A-RPA framework for online adaptive radiotherapy that unifies image synthesis, multimodal registration, contour adaptation and dose prediction within a single framework. The model takes the on-treatment CBCT as the central input and integrates it with available planning information, including planning CT, prior structure contours, prior dose and, when available, auxiliary diagnostic imaging such as MRI or PET-CT. From these inputs, it generates the key outputs required for treatment reassessment and replanning, including synthetic CT, deformable image alignment, updated target and OAR contours, and three-dimensional dose distributions.

The architecture was designed around two complementary components with distinct yet interconnected functions. For CBCT-to-CT synthesis and dose prediction, we used a latent diffusion framework based on Stable Diffusion 3.5 (SD), which enabled generative modeling of anatomically realistic images and dose representations while incorporating radiotherapy-specific context. For anatomical alignment and contour adaptation, we used a DINOv2-based image encoder equipped with lightweight three-dimensional adapters to extract anatomically consistent volumetric features from treatment and planning images. In this design, the diffusion component served as the generative backbone for synthetic image and dose formation, whereas the DINOv2 component served as the representation backbone for registration and segmentation.

The workflow initiates with a diffusion-based synthesis branch, where the daily CBCT is first projected into a latent space via a frozen Stable Diffusion 3.5 (SD) Variational Autoencoder (VAE). These latent representations are stacked into a three-dimensional (3D) volume and processed by a 3D U-shaped Vision Diffusion Transformer (U-ViT). Guided by language embeddings from the SD language encoder, the U-ViT synthesizes high-fidelity CT latent features, which are subsequently decoded to produce an anatomically corrected synthetic CT.

To leverage prior anatomical knowledge, we integrated a secondary pathway centered on a pre-trained DINOv2 foundation model. Planning-phase multi-modal images (CT, MRI, or PET-CT) are encoded into deep semantic features and passed through a 2D-to-3D adapter. This module concatenates the planning features with the sCT features, creating a comprehensive volumetric representation using multiple 3D convolutional layers. This merged feature set drives dual registration and segmentation decoders, which perform deformable alignment of planning priors and update anatomical contours to reflect the patient's current geometry.

Finally, these structure-aware features are ingested by a terminal 3D U-ViT to predict the dose distribution. By decoding the resulting dose features through the SD VAE, the system generates a 3D dose map that accounts for real-time anatomical variations. Notably, the framework maintains the SD VAE in a frozen state, while pre-training the DINOv2 encoder in the pre-training dataset as a robust backbone for medical scene understanding.

In summary, the framework operates predominantly within the latent space, following a sequential pipeline that progresses from a U-ViT for synthesis to a 2D-3D adaptor for feature fusion, and finally to a second U-ViT for dose prediction; within this flow, specialized decoders act as modular branches that are invoked by the agent to output specific synthetic, structural, or dosimetric results as required.

Overall, this One-for-All design integrates generative image modeling, multimodal anatomical alignment, structure-aware contour adaptation and conditional dose generation into a unified online adaptive radiotherapy framework. By coupling these components within a shared architecture and optimizing them jointly on longitudinal clinical data, the model enables coordinated treatment reassessment and replanning across disease sites, imaging modalities and treatment techniques.

\subsection*{Agent orchestration of the A-RPA framework}

We developed an agent-based orchestration framework to integrate the A-RPA model into the iterative decision process of adaptive radiotherapy. The framework was implemented with LangChain and LangGraph\cite{pandya2023}, with GPT-4o\cite{hurst2024} serving as the central multimodal controller for clinician-intent understanding, state tracking, and response generation across the adaptive workflow. For each treatment fraction, the agent received the clinician's query together with available patient-specific information and a pool of candidate outputs generated by the One-for-All model, including original-plan assessment, registered CT, propagated contours, transferred dose, synthetic CT, re-segmented structures and updated dose estimates. Rather than enforcing a fixed computational sequence, the agent interpreted qualitative clinical language and selected the most appropriate result for the current decision stage. Intermediate results and prior interactions were retained in a cache memory, allowing the agent to reuse previously generated outputs, avoid redundant computation and preserve continuity across sequential clinician queries.

After each interaction, the agent updated the case state by integrating the current query, selected results and cached context, and then determined whether the existing information was sufficient for clinical decision-making or whether a more advanced result should be presented to support deeper evaluation. In cases described as showing only slight anatomical change, the agent first prioritized the simplest clinically relevant response, namely whether the original plan could be maintained. When additional review was requested, it could return registration-based results, including the registered CT, propagated contours and transferred dose. When these outputs were considered insufficient, or when the clinician indicated larger anatomical deviation, poor registration quality or the need for full adaptation, the agent escalated to synthesis-based and re-segmentation-based results to support deeper review and replanning.

This design enabled a stepwise, memory-aware and clinically interpretable workflow that more closely reflected online adaptive radiotherapy practice than a rigid automated pipeline. The agent functions as an intelligent assistant that dynamically evaluates each case, intentionally selecting the lowest-computation solution required to validate the day's treatment. For instance, the agent may first perform only an image registration of the planning CT, contours, and dose to the daily CBCT to provide a baseline for the original plan. If a physicist reviews this output and identifies a clinical discrepancy---such as an inaccurate tumor contour---they can trigger the agent to pivot and escalate to the more computationally intensive re-segmentation and dose prediction phases. The agent thus functioned not as an autonomous decision-maker, but as an assistant that anticipated clinical needs, retrieved relevant information, and presented it in a form directly usable for physician review, a paradigm that we term ``human-in-the-loop agentic AI'' for online adaptive radiotherapy.

\subsection*{Domain adaptation of the DINOv2 encoder: pre-training}

To adapt DINOv2 to radiotherapy imaging, we used a parameter-efficient transfer strategy that preserved the broad visual priors learned during large-scale pretraining on natural images while enabling specialization of volumetric medical data and avoiding the risk of overfitting on relatively smaller clinical cohorts. As DINOv2 was originally developed for 2D natural images with a standard grid-based patch embedding, we introduced a learned 3D patch embedding module\cite{xu2025} to transform the synthetic CT volume generated by the synthesis branch into a sequence of volumetric tokens compatible with the pretrained transformer architecture while preserving spatial relationships along all three axes.

The pretrained DINOv2 backbone was kept frozen throughout adaptation, and lightweight trainable 3D adapters\cite{pandya2023} (consisting of compact feedforward networks with residual connections) were inserted to support volumetric radiotherapy representation learning without full backbone finetuning. These adapters functioned as detachable plug-in modules, such that domain-specific adaptation could be introduced or removed without altering the original pretrained encoder weights. The DINOv2 encoder is pre-trained following the same DINOv2 pretraining objectives\cite{oquab2023}.

The adapters were placed in transformer blocks 4, 8 and 24 to capture complementary information across multiple scales: local structural detail (block 4), intermediate anatomical organization (block 8), and higher-level semantic context (block 24). Features from these adapter-enhanced layers were fused into a unified multi-scale volumetric representation for downstream prediction. By restricting optimization to the 3D patch embedding and adapter modules, this strategy enabled efficient transfer of a general-purpose vision foundation model to 3D radiotherapy imaging while preserving backbone stability, reducing computational burden and limiting overfitting on curated clinical cohorts.

\subsection*{Domain adaptation of the encoders and decoders: fine-tuning}

Before construction of the fully integrated One-for-All radiotherapy model, each pretrained component was first adapted to the medical imaging domain through task-specific fine-tuning on publicly available datasets. Because no single public dataset provided unified supervision for synthesis, deformable registration, segmentation and dose prediction, these four modules were optimized separately at this stage, each using its own task-specific data and corresponding supervision. Specifically, the CBCT-to-CT synthesis module was trained with synthesis loss, the registration module with registration loss, the segmentation module with segmentation loss and the dose prediction module with dose loss. For the DINOv2-based modules, this adaptation was performed in a parameter-efficient manner: the pretrained encoder backbone was kept frozen, and only the inserted adapters together with the downstream decoder heads were optimized. This strategy preserved the broad visual priors acquired during large-scale pretraining while enabling domain adaptation to radiotherapy imaging and reducing the risk of overfitting on curated public medical datasets. Collectively, this modular fine-tuning stage allowed each component to acquire task-relevant radiotherapy representations before subsequent integration into the unified framework.

\subsubsection*{Fine-tuning 1: Unsupervised 2D CBCT-to-CT synthesis with text-conditioned latent diffusion}

We developed an unsupervised CBCT-to-CT synthesis framework based on text-conditioned latent diffusion using the publicly available Stable Diffusion 3.5\cite{esser2024} model as a foundation. Rather than predicting CT directly in image space, which is a challenging task given the substantial domain shift between CBCT and CT, the model translated CBCT slices into CT by operating through a shared latent representation. Specifically, CBCT and CT slices were first encoded by a pretrained variational autoencoder into a common latent space, after which synthesis was performed through latent inversion followed by CT-conditioned reverse denoising. This design enabled image translation to be learned in latent space while preserving the powerful generative prior of the pretrained diffusion model, which already captured a rich distribution of natural and medical image structures.

To incorporate radiotherapy-specific context and guide the generative process toward clinically relevant outputs, CBCT and CT slices were jointly modeled with text queries describing both image domain and treatment setting, including treatment site, photon versus proton modality, and conventional versus stereotactic delivery technique. The diffusion backbone was built on the pretrained Stable Diffusion 3.5 2D U-Net, which was not retrained from scratch. Instead, to adapt the model efficiently to radiotherapy imaging, we applied low-rank adaptation (LoRA)\cite{hu2022lora} to the text-conditioning pathway and inserted lightweight trainable adapters into the pretrained diffusion backbone at multiple depths, while keeping the vast majority of original model parameters fixed and frozen. These components were optimized jointly using the standard diffusion denoising objective ($L_{syn}$),

\begin{equation}
L_{syn} = \left\| \widehat{\epsilon} - \epsilon \right\|_{2}^{2}.
\end{equation}

where $\widehat{\epsilon}$ and $\epsilon$ denote the predicted and ground-truth noise, respectively. During inference, each CBCT slice was first encoded into the latent space of the pretrained variational autoencoder. DDIM inversion was then applied to progressively transform the clean CBCT latent $z_{0}^{CBCT}$ into a noisy latent representation $z_{t}^{CBCT}$ under the CBCT condition:

\begin{equation}
z_{t+1}^{CBCT} = \Phi_{CBCT}\!\left(z_{t}^{CBCT}, t\right).
\end{equation}

Here, $\Phi_{CBCT}(\cdot)$ denotes the DDIM inversion update under the CBCT prompts. CT-conditioned reverse denoising was next initialized from this CBCT-derived noisy latent, with the first reverse step explicitly bridging the CBCT and synthetic CT trajectories:

\begin{equation}
z_{t}^{CBCT} \xrightarrow{\ \varphi_{CT}\ } z_{t-1}^{sCT}.
\end{equation}

Here, $\varphi_{CT}(\cdot)$ denotes the DDIM denoising operator under the CT prompts. Subsequent reverse steps then proceeded recursively along the synthetic CT trajectory itself until the final clean synthetic CT latent $z_{0}^{sCT}$ was obtained:

\begin{equation}
z_{k}^{sCT} \xrightarrow{\ \varphi_{CT}\ } z_{k-1}^{sCT}, \qquad k = t-1, \ldots, 1.
\end{equation}

This design allowed the model to retain the anatomical structure encoded in the input CBCT while progressively shifting image appearance toward the CT domain. The final synthetic CT image was generated by decoding $z_{0}^{sCT}$ through the pretrained decoder.

To improve volumetric consistency, inference was performed on overlapping axial slice groups and merged into full three-dimensional volumes. For scans with severe artifacts, normalized metal artifact reduction was optionally applied before latent encoding based on the agent's decision. The resulting synthetic CT volume was subsequently restacked and provided to downstream volumetric modules for registration, segmentation and dose prediction.

\subsubsection*{Fine-tuning 2: Multi-modal deformable registration}

We developed a unified deformable registration framework to map planning CT, T1-weighted MRI, T2-weighted MRI and PET into the synthetic CT space within the One-for-All radiotherapy model. Each training sample consisted of five rigidly registered volumetric inputs: planning CT, T1-weighted MRI, T2-weighted MRI, PET and the synthetic CT produced by the synthesis module.

To connect the synthesis and registration branches, the synthetic CT generated by the synthesis module was used as volumetric input to the adapted DINO encoder. Features from selected adapter-enhanced transformer layers were resized to matched spatial scales using trilinear interpolation, concatenated along the channel dimension and passed to the registration decoder. Using this fused multi-level representation, the decoder predicted modality-specific dense deformation fields to warp each moving modality into the fixed synthetic CT space. The network followed a TransMorph-style\cite{chen2022} design, combining a pretrained DINO-based volumetric encoder with a deformable registration decoder for multimodal alignment.

Training was performed bidirectionally to improve deformation consistency and encourage inverse-consistency in the learned transformations, including forward registration from CT, T1-weighted MRI, T2-weighted MRI and PET to synthetic CT and inverse registration from synthetic CT back to each source modality. It ensures the anatomy is stretched or compressed naturally without non-physical artifacts like ``folding'' or tearing of the tissue. Intuitively, if the model maps a tumor from the planning CT to the daily scan, it must also be able to map it back to its starting position perfectly. This bidirectional constraint acts as a self-correcting mechanism, ensuring that the resulting deformation fields are smooth, invertible, and topologically sound, providing a robust geometric foundation for all subsequent clinical assessments.

The registration objective combined cross-modal image similarity, anatomical mask consistency, groupwise alignment and deformation smoothness regularization. Specifically, cross-modal correspondence was supervised by a modality-independent neighborhood descriptor similarity\cite{heinrich2012} term $L_{sim}$, anatomical overlap within the shared body region by a body-mask consistency term $L_{mask}$, agreement among all warped modalities in the synthetic CT space by a groupwise consistency\cite{zhang2021} term $L_{group}$, and transformation regularity by a smoothness penalty $L_{smooth}$. The total registration loss $L_{reg}$ was defined as

\begin{equation}
L_{reg} = L_{sim} + \lambda_{mask} L_{mask} + \lambda_{group} L_{group} + \lambda_{smooth} L_{smooth}
\end{equation}

with $\lambda_{mask} = 1$, $\lambda_{group} = 0.4$, and $\lambda_{smooth} = 0.15$. During inference, the predicted low-resolution deformation fields were upsampled to full resolution and applied to the original images, contours and prior dose distributions for downstream tasks.

\subsubsection*{Fine-tuning 3: Structure segmentation with planning-aware contour adaptation}

We developed a planning-aware three-dimensional segmentation framework to delineate OAR and target volumes on treatment-day images by adapting prior planning contours to treatment-specific anatomy. The model was designed as a unified multi-site system for head-and-neck, thoracic, abdominal and pelvic radiotherapy. To support learning across anatomically heterogeneous settings, all site-specific annotations were remapped into a shared 76-channel label space containing background, body, normal OARs and target volumes. The framework used a single three-dimensional segmentation network to jointly estimate deformable alignment and updated structure masks from current image appearance and patient-specific planning priors.

For each case, the network took as input the target synthetic CT, a rigid-aligned prior CT and the corresponding rigid-aligned prior structure masks. The model then predicted both a dense deformation field and segmentation logits in the target image space. The deformation field was used to warp prior anatomical and contour information toward the target anatomy, allowing segmentation to be guided by both image evidence and propagated planning context. To further preserve continuity of target volumes during adaptation, the warped prior target masks were added to the corresponding target-related logit channels before final prediction.

Training was performed in a bidirectional manner by randomly swapping the roles of the current image and the affine planning image. In one training instance, the current image served as the target and the affine planning image served as the prior; in another, the reverse assignment was used. This symmetric design encouraged the model to learn anatomically consistent contour adaptation between treatment and planning domains without restricting learning to a single propagation direction.

The overall objective $L_{seg}$ combined segmentation supervision with deformation-guided image consistency:

\begin{equation}
L_{seg} = L_{Dice}^{sig} + L_{Dice}^{soft} + \lambda_{reg} \left\| \phi\!\left(I_{rigid}\right) - I_{cur} \right\|_{1}.
\end{equation}

Here, $\phi$ is the registration field estimated by the network, $L_{Dice}^{sig}$ denotes sigmoid-based Dice loss applied to the tumor-related channels, $L_{Dice}^{soft}$ denotes softmax-based Dice loss applied to the remaining non-tumor organs, and the image-consistency term penalizes disagreement between the warped prior image and the target image under the predicted deformation field. In our implementation, $\lambda_{reg} = 1$. During inference, the network produced the deformation field and final structure logits in a single forward pass, and overlapping patch predictions were merged to generate full-volume contours.

\subsubsection*{Fine-tuning 4: Dose prediction with anatomy-, structure-, and metadata-conditioned latent diffusion}

We developed a conditional latent diffusion module for treatment-specific three-dimensional dose prediction in online adaptive radiotherapy. This module was built on a Stable Diffusion 3.5-style latent diffusion architecture, but was adapted for volumetric dose generation rather than two-dimensional image synthesis. In the overall framework, the synthetic CT branch used an SD3.5-based 2D latent diffusion model to translate daily CBCT slices into synthetic CT slices, which were then reassembled into a three-dimensional anatomical volume. By contrast, the dose branch used an SD3.5-based latent generative design coupled with a dedicated 3D denoising U-Net, enabling direct generation of the full three-dimensional dose representation in latent space.

Rather than predicting dose from the treatment-day image alone, this module was conditioned on the anatomical and structural information produced by the upstream framework. Specifically, the treatment-day three-dimensional image, together with aligned planning information and updated target and OAR contours from the registration and segmentation branches, was encoded as the principal spatial context for dose generations. Prior dose and radiotherapy metadata were incorporated as additional conditioning signals. A language-conditioning pathway was used to embed the treatment-specific metadata, including anatomical site, treatment modality and delivery technique, so that the model could incorporate both anatomical structure and clinical planning context.

The final conditioning representation therefore integrated treatment-day anatomy, propagated and updated in prior modules, for structural guidance, prior dose context, and treatment-specific metadata embeddings. The model was trained using a standard diffusion denoising objective on the target dose latent ($L_{dose}$),

\begin{equation}
\mathcal{L}_{dose} = \left\| \widehat{\epsilon}_{dose} - \epsilon_{dose} \right\|_{2}^{2},
\end{equation}

where $\widehat{\epsilon}$ and $\epsilon$ denote the predicted and sampled noise for dose distribution, respectively. To reduce overreliance on imperfect historical dose estimates and improve robustness during adaptation, we introduced prior-dose dropout during training. During inference, the model generated the full three-dimensional dose latent, which was then decoded into voxel space to produce the final predicted dose distribution for downstream dosimetric evaluation.

\subsection*{Joint post-training of the unified One-for-All model on the internal clinical cohort}

After task-specific adaptation on public datasets, we jointly post-trained the full A-RPA framework on the internal clinical cohort from Emory University Hospital. In contrast to the public datasets, the internal cohort contained complete radiotherapy information within the same clinical cases, including planning CT, serial treatment-time CBCT, prior structure contours, dose distributions and, when available, MRI and PET--CT. This comprehensive longitudinal dataset enabled all components of the framework to be connected within a single adaptive radiotherapy setting and optimized jointly.

At this stage, synthesis, registration, segmentation and dose prediction were trained as a cascaded system rather than as isolated tasks. The synthesis module provided anatomically enhanced image information for downstream analysis, the registration module aligned planning and auxiliary imaging to the treatment CBCT space, the segmentation module updated OAR and target contours, and the dose module predicted the treatment-specific three-dimensional dose distribution from the resulting anatomical and structural context. Because these inputs and labels coexisted within the same treatment courses, upstream modules could be optimized not only for their own intermediate objectives, but also for their contribution to downstream anatomical and dosimetric prediction.

To reflect this dependency structure, supervision was applied hierarchically at the module level. The synthesis module was optimized with

\begin{equation}
L_{syn}^{joint} = \lambda_{1}\left(L_{syn} + L_{reg} + L_{seg} + L_{dose}\right),
\end{equation}

the registration module with

\begin{equation}
L_{reg}^{joint} = \lambda_{2}\left(L_{reg} + L_{seg} + L_{dose}\right),
\end{equation}

the segmentation module with

\begin{equation}
L_{seg}^{joint} = \lambda_{3}\left(L_{seg} + L_{dose}\right),
\end{equation}

and the dose module with

\begin{equation}
L_{dose}^{joint} = \lambda_{4} L_{dose}.
\end{equation}

The overall joint optimization objective was therefore

\begin{equation}
L_{joint} = L_{syn}^{joint} + L_{reg}^{joint} + L_{seg}^{joint} + L_{dose}^{joint}.
\end{equation}

Here, $L_{syn}$, $L_{reg}$, $L_{seg}$ and $L_{dose}$ denote the synthesis, registration, segmentation and dose objectives, respectively, and $\lambda_{1} = 0.4$, $\lambda_{2} = 0.4$, $\lambda_{3} = 0.7$ and $\lambda_{4} = 1$ are empirical weighting factors. This progressive supervision strategy encouraged each upstream component to generate outputs that were not only locally accurate, but also maximally informative for subsequent tasks.

\bibliographystyle{unsrt}
\bibliography{references}

\end{document}